\documentclass[runningheads]{llncs}
\usepackage[T1]{fontenc}
\usepackage{graphicx}
\usepackage{cite}
\usepackage{amsmath,amssymb,amsfonts}
\usepackage{algorithmic}
\usepackage{graphicx}
\usepackage{textcomp}
\usepackage{xcolor}
\usepackage{graphicx}
\usepackage{algorithm}
\usepackage{algorithmic}
\usepackage{amsmath,amssymb,amsfonts}
\usepackage{textcomp}
\usepackage{xcolor}
\usepackage{booktabs}
\usepackage{multirow}
\usepackage{booktabs}
\usepackage{threeparttable}
\usepackage{newfloat}
\usepackage{listings}
\usepackage{url}
\usepackage{tabularx}

\begin{document}
\title{Logic Error Localization in Student Programming Assignments Using Pseudocode and Graph Neural Networks}
%
%

\author{Zhenyu Xu\textsuperscript{1} \and
Kun Zhang\textsuperscript{2} \and
Victor S. Sheng\textsuperscript{1}}
\authorrunning{Z. Xu et al.}
%
\institute{Department of Computer Science, Texas Tech University, TX, USA\\
\email{\{zhenxu, victor.sheng\}@ttu.edu} \and
Department of Computer Science, Xavier University of Louisiana, LA, USA\\
\email{kzhang@xula.edu}
}

\maketitle              

\begin{abstract}
Pseudocode is extensively used in introductory programming courses to instruct computer science students in algorithm design, utilizing natural language to define algorithmic behaviors. This learning approach enables students to convert pseudocode into source code and execute it to verify their algorithms' correctness. This process typically introduces two types of errors: syntax errors and logic errors. Syntax errors are often accompanied by compiler feedback, which helps students identify incorrect lines. In contrast, logic errors are more challenging because they do not trigger compiler errors and lack immediate diagnostic feedback, making them harder to detect and correct. To address this challenge, we developed a system designed to localize logic errors within student programming assignments at the line level. Our approach utilizes pseudocode as a scaffold to build a code-pseudocode graph, connecting symbols from the source code to their pseudocode counterparts. We then employ a graph neural network to both localize and suggest corrections for logic errors. Additionally, we have devised a method to efficiently gather logic-error-prone programs during the syntax error correction process and compile these into a dataset that includes single and multiple line logic errors, complete with indices of the erroneous lines. Our experimental results are promising, demonstrating a localization accuracy of 99.2\% for logic errors within the top-10 suspected lines, highlighting the effectiveness of our approach in enhancing students' coding proficiency and error correction skills.

\end{abstract}

\section{Introduction}


Pseudocode is commonly used in introductory programming courses to teach algorithms to computer science students \cite{dirgahayu2017automatic}. It is defined using natural language familiar to the students, conveying the behavior of algorithms. For students with different linguistic backgrounds, pseudocode can be written in their native languages, with the high level of abstraction allowing for translation into English while preserving the original meaning. This approach lowers the language barrier students might face when learning to define algorithms. Pseudocode can include mathematical expressions when they simplify the description of certain behaviors within an algorithm. It allows students to read and understand algorithms by abstracting away the details of programming languages, enabling them to focus on defining the algorithm to solve problems, rather than its technical implementation in code.

During the process where students learn to convert pseudocode into source code, they frequently encounter syntax errors and logic errors, also known as semantic errors. For syntax errors, the compiler can report error messages including suspicious syntax error tokens and syntax error types, which can be used as clues to track and determine syntax errors for students. For example, Microsoft Visual C++ compiler, GCC (the GNU compiler collection), and Clang/LLVM all can catch syntax errors and memory errors in programs \cite{Stanier2013-dj}. 

Unlike syntax errors, logic errors do not trigger compiler error reports, are more challenging to localize in student programming assignments. Logic errors can be caused by misunderstanding program specifications or by minor mistakes in the code, such as a wrong iteration number in a loop or a misplaced decimal point. These errors can result in programs to fail test cases and can be challenging for students to locate without any guidance from compilers. Even experienced instructor can spend a considerable amount of time finding these errors, as understanding the program specification and logical structure of the code is an essential step in correcting logic errors. Common types of logic errors are given in Table 1. 

\begin{table}[!ht]
    \caption{Common types of logic errors and their descriptions}
    \label{table1}  
    \centering
    \renewcommand{\arraystretch}{1.5}  
    \begin{tabularx}{\textwidth}{|>{\hsize=.6\hsize\linewidth=\hsize}X|>{\hsize=1.4\hsize\linewidth=\hsize}X|}
    \hline
    \textbf{Type of Logic Error} & \textbf{Description} \\ 
    \hline
    Loop Condition & Incorrect iteration numbers, inequality, or logical conjunctions in the for/while loops. \\
    \hline
    Condition Branch & Incorrect logical expressions in the if condition. \\
    \hline
    Statement Integrity & Statement lacks a self-consistent logical structure after the condition. \\
    \hline
    Variable Initialization & Incorrect declaration and initialization of variables. \\
    \hline
    Data Type & Incorrect data type. \\
    \hline
    Computation & Incorrect basic math symbols or missing mathematical brackets. \\
    \hline
    \end{tabularx}
\end{table}
The difficulty of identifying logic errors varies by type. Loop condition and data type errors are usually simpler to spot and fix due to their structured and predictable patterns, such as the for keyword in "for (init; condition; increment)" often located on a single line. This clarity and consistency in structure make it easier for automated tools to detect and propose corrections. In contrast, errors in condition branch, such as those involving multiple "if" and "else if" statements over several lines, present more challenges. Their complexity and the variety of conditions and logic across different branches make automated correction harder. This is because accurately fixing such errors often requires an understanding of the program's broader logic, beyond just the immediate context of the errors. Each logic error type has unique characteristics that influence the ease of identification and repair.

Since compilers report error lines for syntax errors, students usually do not face significant challenges in locating them independently. Therefore, our focus is on addressing the challenge of localizing logic errors at the line level. We propose a novel approach to enhance logic error localization in introductory programming courses, drawing inspiration from the DrRepair \cite{yasunaga2020graph} model. Our method combines the analysis of source code and pseudocode, utilizing bidirectional Long Short-Term Memory (BiLSTM) \cite{graves2012long} networks and a graph attention layer \cite{velickovic2017graph} to decipher code structure and logic. Furthermore, we employ CodeBERT \cite{feng2020codebert} to evaluate the semantic similarity between code and pseudocode, allowing the model to predict error probabilities in code lines while adjusting its focus based on semantic alignment. Throughout training, we dynamically balance the emphasis between error prediction accuracy and semantic understanding, aiming to improve logic error detection by integrating advanced techniques and adaptive learning strategies.

In our study, we explore a method to collect logic errors during the syntax error repair process of DrRepair, utilizing the SPoC dataset \cite{Kulal2019-eo} composed of C++ programs from programming competitions. DrRepair, originally designed to fix syntax errors, iteratively repairs programs until they pass all test cases. However, this process often results in programs with logic errors. By analyzing these iterations, we construct a dataset containing various types of logic errors, providing a valuable resource for studying logic error localization and correction. Furthermore, we conducted a comparison of our approach with existing state-of-the-art tools for logic error localization and analyzed the impact of different types of logic errors on localization accuracy.

Our contributions are listed as follows:
\begin{enumerate}
    \item We introduce a novel technique that leverages pseudocode and employs semantic alignment to assist in localizing logic errors in students' programs.
    \item We utilize a graph-based approach in both source code and pseudocode to enhance the localization of logic errors.
    \item We create a dataset specifically tailored for logic error analysis, providing a valuable resource for further research in this area.
\end{enumerate}

\section{Related Work}
In this section, we review the existing literature on automated program repair with deep learning, logic error localization, and graph neural networks, which form the foundation for our approach.
\subsection{Automated Program Repair with Deep Learning}

Deep learning has significantly advanced the field of program repair, particularly in syntax error correction. Gupta et al. introduced DeepFix, a sequence-to-sequence model that repairs syntax errors in C programs but does not consider the program's structure \cite{gupta2017deepfix}. To capture this structure, Graph Neural Networks (GNNs) have become popular. Allamanis et al. used Gated Graph Neural Networks to represent both syntactic and semantic aspects of code \cite{allamanis2017learning}. Dinella et al. proposed Hoppity, which transforms a buggy program into a graph to predict error locations and their repairs \cite{dinella2020hoppity}. Yasunaga et al. designed a program-feedback graph, combining source code and compiler feedback to improve syntax error repair using GNNs \cite{yasunaga2020graph}. Chen et al. introduced PLUR, an algorithm that simplifies program learning and repairing through a program-feedback graph \cite{chen2021plur}. Li et al. employed context learning and tree transformation to fix syntax errors requiring consecutive changes \cite{li2022dear}. These advancements demonstrate the efficacy of combining deep learning with program structures and compiler feedback for syntax error correction.

\subsection{Logic Error Localization}

In student programming assignments, detecting and fixing logic errors can be achieved using several approaches. Test cases are commonly used to verify if the program's output matches the expected results, helping to identify any discrepancies. Static analysis tools are employed to examine the code without executing it, pinpointing potential logic issues \cite{StaticAnalysis}. Additionally, Automated Program Repair tools like Tarantula \cite{Jones2005Tarantula}, Ochiai \cite{Abreu2007Ochiai}, and DStar \cite{Wong2014DStar} can suggest corrections at specific lines in the code, using various techniques to locate and address errors effectively. Raana et al. proposed a system to detect logic errors in C++ codes, extract the dependency of methods or functions among source codes, and classify detected logic errors based on a decision tree \cite{raana2016c++}.  Lee et al. presented FixML, a system for automatically generating feedback on logic errors for students’ programming assignments \cite{Lee2018-ib}. Yoshizawa et al. proposed a logic error detection system based on program structure patterns \cite{BibitemrefArticle_Yoshizawa2019-pm}. Rahman et al. applied a language model to evaluate source codes using a bidirectional long short-term memory (BiLSTM) neural network \cite{BibitemrefArticle_Rahman2021-bq}. Matsumoto et al. provide an iterative trial model to repair multiple logic errors in source codes \cite{matsumoto2021model}. NeuralBugLocator (NBL)  \cite{gupta2019neural} is a deep learning-based technique developed by Gupta et al. for localizing logic errors in student programs with respect to a failing test without executing the program. Fine-grained Fault Localization \cite{nguyen2022ffl} combines syntactic and semantic analysis to more accurately localize errors in student programs, outperforming existing techniques on the Prutor and Codeflaws datasets.

\subsection{Graph Neural Network}

Graph Neural Networks have emerged as a powerful tool for learning representations of graph-structured data, enabling a wide range of applications across various domains. Scarselli et al. introduced the concept of GNNs, proposing a model that extends traditional neural networks to handle graph data by leveraging the graph structure to propagate node features \cite{scarselli2009}. Veličković et al. introduced the Graph Attention Network (GAT), which incorporates attention mechanisms to weigh the importance of neighboring nodes during feature aggregation, allowing for more flexible and powerful graph representations \cite{velickovic2017}. Olah and Perez demonstrated the application of GNNs in predicting traffic flow, showcasing their ability to capture spatial and temporal dependencies in complex systems. Monti et al. highlighted the use of GNNs for detecting fake news on social media platforms by modeling the relational information among users and news articles \cite{monti2019}. Stokes et al. showcased the application of deep learning, including GNNs, in discovering new antibiotics, demonstrating the potential of GNNs in drug discovery and biomedical research \cite{stokes2020}. Sanchez-Gonzalez et al. presented a graph network-based approach for simulating complex physical systems, illustrating the versatility of GNNs in modeling physical interactions \cite{Sanchez-Gonzalez2020}. We employ the attention mechanism from GATs to update the embeddings of nodes in the graph. Specifically, each node updates its representation by considering its connections to other nodes, including those in the source code and pseudo-code, and their corresponding weights.

\section{Approach}
In this section, we detail our proposed method for logic error localization, including the model architecture, the construction of our logic error dataset, and the experimental setup.
\subsection{Model Architecture}

The model architecture shown in Figure 1 is an encoder framework based on DrRepair, to identify logic errors in code. It processes both the source code and corresponding pseudocode to encode the information, then outputs predictions for the indices of erroneous lines. 

\begin{figure}[!htbp]
\centerline{\includegraphics[scale=0.17]{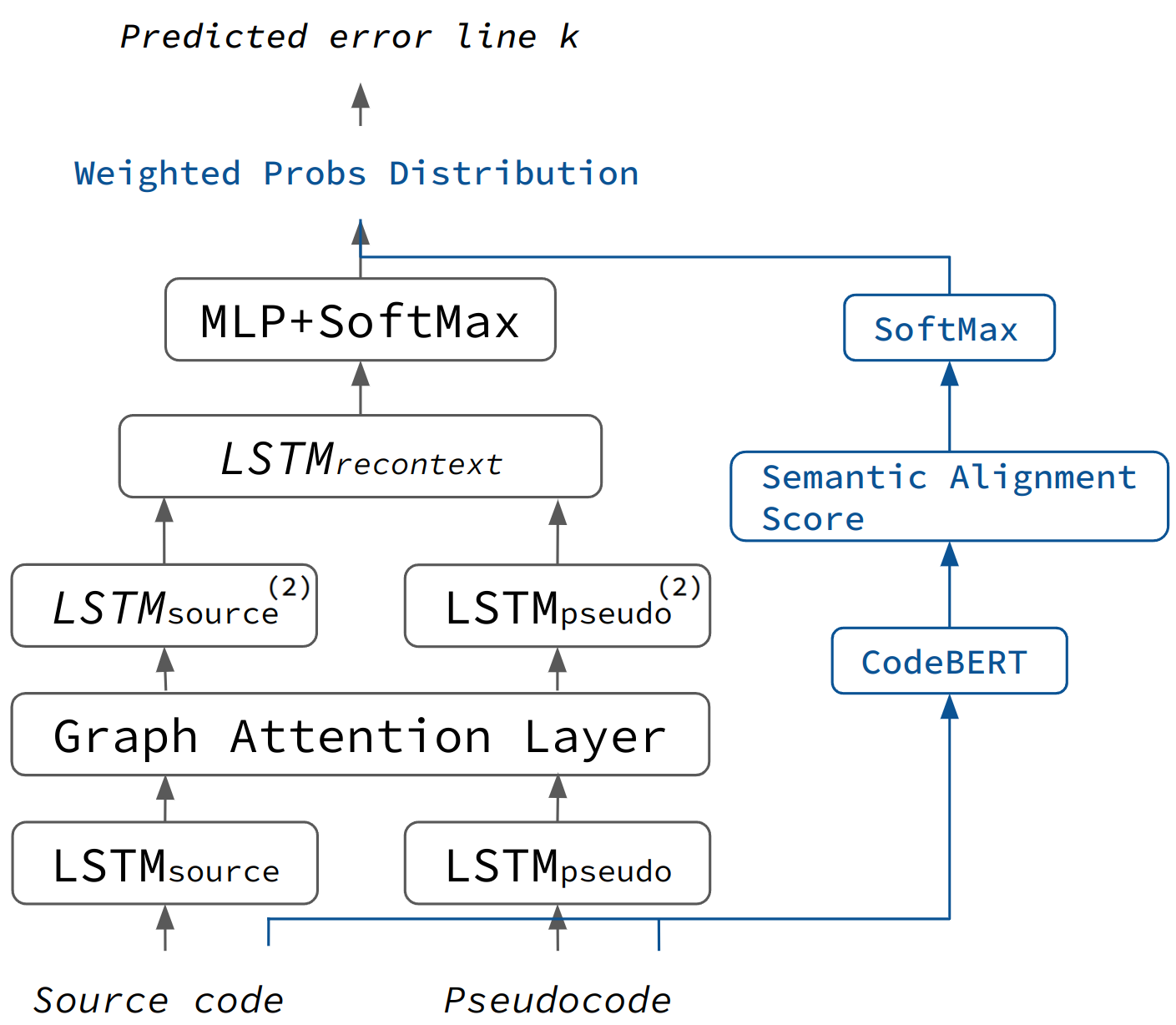}}
\caption{Model architecture.}
\label{p3}
\end{figure}

\subsubsection{Model Overview}

Initially, lines of source code and pseudocode are each processed by corresponding bidirectional LSTM networks, \( LSTM_{source} \) and \( LSTM_{pseudo} \), to generate a hidden state \( h \) for every line. Following this, a graph attention layer, denoted as \( g = Graph(h) \), leverages the structural connections within the code to facilitate information flow and enhance these hidden states. Concurrently, CodeBERT is employed to compute Semantic Alignment Scores, assessing the semantic similarity between the source code and pseudocode across equivalent lines. The model further processes these states with another layer of \( LSTM_{source}^{(2)} \) and \( LSTM_{pseudo}^{(2)} \), enriching the hidden state for each line. Through a re-contextualization function \( x = context(g) \), these states are amalgamated into a cohesive line embedding \( s_i \), transitioning the representation from a token-based to a line-based level, thereby refining the model’s predictive capability. In the final step, the model utilizes an MLP followed by a SoftMax layer to deduce the probability of errors across the code lines. This phase incorporates the Semantic Alignment Scores into the model's error prediction, creating a weighted probability distribution. Specifically, if a line is initially deemed likely to contain an error but exhibits a high semantic alignment score, its probability of being the erroneous line is adjusted accordingly.

The training of the model involves dynamic adjustments of the weights assigned to Semantic Alignment Loss. Initially, the training concentrates on reducing the Cross-Entropy Loss to enhance the model's ability to accurately predict logical error lines. As the training progresses and the model's proficiency in identifying error lines increases, the emphasis gradually shifts towards the Semantic Alignment Loss, aiming to foster a deeper semantic understanding and alignment between the source code and pseudocode.

\subsubsection{Graph Attention Layer}
The graph attention layer forms a key part of our model, creating connections between tokens from the source code and pseudocode that are essential for identifying logic errors. Through a graph $G = (V, E)$, with nodes $V$ representing tokens and edges $E$ connecting matching tokens, the model captures semantic relationships within the code. This approach, advancing beyond DrRepair's framework, leverages pseudocode to enhance logic error localization.

In this layer, token representations are transformed and analyzed to ascertain the importance of each token's connections. Equations (1) through (4) detail this process, which involves linear transformation, attention score computation, normalization through SoftMax, and the aggregation of neighboring node information. This mechanism assigns weights to nodes, guiding the model's focus to the most significant tokens for error correction. The utilization of graph attention, as defined by Velicković et al. \cite{velickovic2017graph}, empowers the model to trace and emphasize tokens crucial for understanding the logical flow of code.

\begin{align}
z_i^{(l)}&=W^{(l)}h_i^{(l)}\\
e_{ij}^{(l)}&=\text{LeakyReLU}(\vec a^{(l)^T}(z_i^{(l)}||z_j^{(l)}))\\
\alpha_{ij}^{(l)}&=\frac{\exp(e_{ij}^{(l)})}{\sum_{k\in \mathcal{N}(i)}^{}\exp(e_{ik}^{(l)})}\\
h_i^{(l+1)}&=\sigma\left(\sum_{j\in \mathcal{N}(i)} {\alpha^{(l)}_{ij} z^{(l)}_j }\right)
\end{align}

\subsubsection{Logic Error Line Prediction with Semantic Alignment}

Employing CodeBERT, we derive semantic alignment scores between the source code lines $x_{1:L}$ and their equivalent pseudocode lines $y_{1:L}$ to evaluate their semantic congruence:
\begin{align}
\alpha_{1:L} &= \text{Softmax}(\text{CodeBERT}(x_{1:L}, y_{1:L}))
\end{align}

Table 2 showcases a side-by-side comparison of pseudocode and corresponding source code. Throughout the training phase, the model dynamically adjusts the weights assigned to these semantic alignment scores to fine-tune learning priorities. Initially, the emphasis is placed on minimizing Cross-Entropy Loss to expedite the enhancement of error line prediction accuracy. As the model's predictive proficiency evolves, we progressively accentuate the semantic alignment scores to bolster the model's semantic comprehension and correlation:
\begin{align}
\tilde{p}(k|x_{1:L}, \alpha_{1:L}) &= \text{Softmax}(\alpha_{1:L} \odot \text{MLP}(x_{1:L}))
\end{align}
where $\alpha_{1:L}$ is adaptively modified over the training period to gradually shift focus towards semantic alignment.

The objective of the training regimen is to minimize a holistic loss that amalgamates the Cross-Entropy Loss, ensuring precise error line detection, with a component that assesses the degree of semantic alignment. This dual-focus strategy endeavors to balance the accurate identification of errors while preserving semantic consistency between the source code and pseudocode. The augmented loss function is delineated as:
\begin{align}
\mathcal{L}_{CE} &= -\sum_{i=1}^{L} y_i \log(p(k=i|x_{1:L}))
\end{align}

Here, $\mathcal{L}_{CE}$ represents the Cross-Entropy Loss, with $y$ indicating the true labels of the error lines, ensuring that the model not only identifies error lines with heightened accuracy but also deepens its semantic analysis, fostering an enriched understanding of the logical relationships encoded in the pseudocode.


\begin{table}[htbp]
\centering
\caption{Example of pseudocode and corresponding source code.}
\label{tab:example}
\begin{tabularx}{\textwidth}{|c|>{\ttfamily}X|>{\ttfamily}X|}
\hline
\textbf{i} & \multicolumn{1}{c|}{\textbf{Pseudocode \( x_i \)}} & \multicolumn{1}{c|}{\textbf{Source Code \( y_i \)}} \\ \hline
0 & s = string & string s; \\
1 & len = integer & int len; \\
2 & let k, ans be integer & int k, ans = 0; \\
3 & read a & cin >> s; \\
4 & set len to size of s & len = s.size(); \\
5 & for i = 0 to len exclusive & for (int i = 0; i < len; i++) \{ \\
6 & for j = 1 to len exclusive & \ \ for (int j = 1; j < len; j++) \{ \\
7 & for k = 0 to infinity & \ \ \ for (k = 0; k < len; k++) \{ \\
8 & if i + k is greater than len or j + k is greater than len or s[i + k] is not equal to s[j + k], break & \ \ \ \ if ((i + k > len) || (j + k > len) || (s[i + k] != s[j + k])) break;\\
9 & & \ \ \ \} \\
10 & set ans to max of ans, k & \ \ ans = max(ans, k); \\
11 & & \ \} \\
12 & print ans & cout << ans << endl; \\
13 & & return 0; \\
14 & & \} \\
\hline
\end{tabularx}
\end{table}

\subsection{Logic Error Dataset Construction}

\subsubsection{SPoC Dataset}
The SPoC data consists of 18,356 C++ programs, which are collected from programming competitions. Each program has its own human-written pseudocode, and its public and hidden test cases. Kulal et al. \cite{Kulal2019-eo} generate a functional correct program from its corresponding pseudocode in SPoC. Each pseudocode line can be translated to a code line, which can have multiple candidate translations. A program can be synthesized by choosing a suitable candidate translation for each pseudocode line. The program is evaluated against both its public and hidden test cases in SPoC. For each program, there are 5 to 10 public test cases to ensure that the program passes its preliminary functional tests.

\subsubsection{Construction Procedure}
We collect emerging logic errors during DrRepair's syntax error repair process and construct datasets containing programs with logic errors, based on the SPoC dataset. Our observations reveal that DrRepair's repair process on SPoC data generates a significant number of logic errors, encompassing various types of logic errors. DrRepair aims to correct programs with syntax errors, ensuring they pass all test cases. It begins by identifying and fixing syntax errors in an initial program, a process that may require multiple iterations. If the program remains incorrect after an iteration, DrRepair continues to predict and fix errors. Each iteration produces a repaired program by replacing the predicted error line with a modified code line. Successful correction is achieved when a repaired program passes all test cases. However, DrRepair may fail if the repair attempts exceed a set limit (e.g., 100 attempts) or if no suitable code candidates are available to correct the last syntax error in the current corrected program.

DrRepair relies on error messages as hints and cannot directly repair logic errors. During the repair process, if a program is corrected but still contains logic errors, indicating it cannot pass the test cases, DrRepair will backtrack to the previous step and attempt another candidate of correction code lines. This scenario may occur multiple times throughout the repair process, resulting in programs with logic errors being generated. To identify the exact lines with logic errors, we compare programs that still have logic errors after attempted fixes with those that have successfully passed the test cases. By deliberately replacing these identified lines with logic errors, we can artificially create programs that contain these errors for further testing and analysis.

\section{EXPERIMENTS}
In this section, we outline the research questions, experimental methodology, and results of our study.
\subsection{Research Questions}
\subsubsection{RQ1: Performance Comparison with State-of-the-Art Tools}
How does the performance of our model compare with other state-of-the-art tools for logic error localization?

\subsubsection{RQ2: Impact of Logic Error Types}
What is the impact of different types of logic errors on the accuracy of error localization? Do certain types of logic errors pose greater challenges?

\subsection{Experimental Methodology}
\subsubsection{Datasets}

We have curated two specialized datasets, S-Logic-Err and M-Logic-Err, designed to evaluate programs that pass some but fail other test cases. Programs that fail all tests often need extensive rewriting, which reduces the relevance of fault localization. The S-Logic-Err dataset focuses on single-line code errors, while M-Logic-Err covers multi-line errors, thus addressing a wider spectrum of bug localization complexities. Each dataset includes over 500 unique programming challenges, with S-Logic-Err containing approximately 3800 programs and M-Logic-Err about 1500 programs. On average, each program is about 30 lines long. To rigorously evaluate and fine-tune our models, we utilize a five-fold cross-validation method. Each dataset entry is structured with a program ID, source code, pseudocode, and the index of the logic error line. This format provides a robust framework for analyzing and testing various fault localization strategies, facilitating comprehensive studies on the efficacy of different methods. 

\subsubsection{Types of logic errors and their examples.} In the Logic-Err dataset we designed, we emphasize the variety of logic errors to reflect real-world coding issues. Table 3 offers examples for each error type, showcasing both the incorrect and the corrected code lines.

\begin{table}[!ht]
\caption{Examples of common types of logic errors in our dataset, including logic error lines and their correct lines.}
\label{table:logic_errors}
\centering
\renewcommand{\arraystretch}{1.5}  

\begin{tabularx}{\textwidth}{
  |>{\hsize=0.6\hsize\linewidth=\hsize\ttfamily}X|
  >{\hsize=1.2\hsize\linewidth=\hsize\ttfamily}X|
  >{\hsize=1.2\hsize\linewidth=\hsize\ttfamily}X|
}
\hline
\textbf{Type of Logic Error} & \textbf{Logic Error Line} & \textbf{Correct Line} \\ 
\hline
Loop condition & for (int i = 1; i < i; i++) & for (i = 1; i < 10; i++) \\ 
\hline
Condition branch & if (n >= 1) & if (n <= 1) \\ 
\hline
Statement integrity & for (i = 1; i <= 10; i++) \{ sum = i; printf(sum);\}  & for (i = 1; i <= 10; i++) \{ sum += i; printf(sum);\} \\ 
\hline
Variable initialization & int t = red = green = blue = 29; & int t = 29, red, green, blue; \\ 
\hline
Data type & long long n, m, x & int n, m, x \\ 
\hline
Computation & int mid = low + high / 2; & int mid = (low + high) / 2; \\ 
\hline
\end{tabularx}
\end{table}

\subsubsection{Baselines}
In our study, we compare our method with three established tools for finding errors in code: Tarantula, Ochiai, and DStar. They are spectrum-based fault localization (SBFL) techniques that are primarily used to identify fault locations in software by analyzing the execution traces of passing and failing test cases. They compute suspiciousness scores for each program element (like lines of code or blocks) based on how frequently these elements are executed in passing versus failing test runs. Tarantula checks how each part of the program acts in tests that pass and tests that fail. It uses colors to show how likely it is that each part has an error, helping developers find mistakes more quickly. Ochiai works similarly to Tarantula but uses a specific mathematical formula to decide how suspicious each part of the program is. It considers how often each part appears in passing and failing tests to figure out its connection to errors. DStar also looks for errors in the program but uses a different formula that can be adjusted to better suit different situations.

\subsubsection{Metric}
We assess our model's performance using localization accuracy, which focuses on accurately identifying the lines where logic errors occur. For evaluation, we measure how effectively the model localizes errors within the top results of our rankings, specifically reporting the accuracy at the top-1, top-5, and top-10 positions.

\subsubsection{Implementation Details}
In the dataset construction, we increase attempt limit to 300 to generate more candidates containing logic errors. We employ semantic alignment by utilizing the Semantic Textual Similarity (STS) task with CodeBERT. This involves comparing two pieces of text, typically code snippets or a combination of code and its description, to generate a similarity score. For our specific application, we use a version of CodeBERT that has been finetuned on a C++ code corpus, available at \url{https://huggingface.co/neulab/codebert-cpp}.

\subsection{Experimental Results}
In this section, we present the findings from our experiments, focusing on the performance comparison with state-of-the-art tools and the impact of different logic error types on localization accuracy.

\subsubsection{RQ1: Performance Comparison}
To assess the effectiveness of different methods in localizing logic errors, we conducted comprehensive experiments across two distinct datasets, S-Logic-Err and M-Logic-Err. These datasets are designed to challenge the models with single-line and multi-line logic errors, respectively. The results of these experiments are summarized in Table 4, which presents the localization accuracy at the top-1, top-5, and top-10 ranks.

\begin{table*}[!ht] 
\centering
\caption{Logic error localization results on the S-Logic-Err and M-Logic-Err datasets, presented in terms of top-n (\%) accuracy.}
\label{table4}
\begin{tabular}{ccccccc}
    \toprule
    & \multicolumn{3}{c}{S-Logic-Err Dataset} & \multicolumn{3}{c}{M-Logic-Err Dataset} \\
    \cmidrule(lr){2-4} \cmidrule(lr){5-7}
    Method & Top-1 & Top-5 & Top-10 & Top-1 & Top-5 & Top-10 \\
    \midrule
    Tarantula & 18.6 & 41.5 & 62 & 11.6 & 35.1 & 55.7 \\
    Ochiai & 20.4 & 52.2 & 78.9 & 18.7 & 46.2 & 79.4 \\
    DStar & 29.7 & 58.4 & 78.3 & 22.6 & 42.8 & 71.5 \\
    Our Method & 36.1 & 71.2 & 99.2 & 28.6 & 68.3 & 96.4 \\
    \bottomrule
\end{tabular}
\end{table*}

The results indicate that our method outperforms traditional fault localization techniques such as Tarantula, Ochiai, and DStar, particularly in higher recall scenarios (Top-5 and Top-10). This suggests that our approach, which integrates advanced semantic analysis, provides more accurate and reliable localizations across different types of logic errors. We observed a consistent improvement in performance across both datasets.

\subsubsection{RQ2: Impact of Logic Error Types}

To understand the impact of different logic error types on fault localization effectiveness, we analyzed our method's performance using the Top-1 localization ratio. Table 5 presents the distribution of logic errors by type, their proportion in our datasets (S-Logic-Err and M-Logic-Err), and the localization success rate for each type.

\begin{table}[!ht] 
\centering
\caption{Distribution of logic error types and the localization ratio of our method on two datasets (S-Logic-Err and M-Logic-Err)}
\label{table5}
\begin{tabular}{ccccccc}    
\toprule    
\multicolumn{1}{c}{} & \multicolumn{3}{c}{S-Logic-Err Dataset}&\multicolumn{3}{c}{M-Logic-Err Dataset}\\  
\cmidrule(lr){2-4} \cmidrule(lr){5-7}
Type of Logic Error & Total & Proportion & Loc Ratio & Total & Proportion & Loc Ratio \\
\midrule
Loop Condition & 844 & 21.8\% & 63.2\% & 1304 & 37.3\% & 28.9\% \\   
Condition Branch & 906 & 23.4\% & 26.1\% & 1231 & 35.2\% & 24.9\% \\ 
Statement Integrity & 1212 & 31.3\% & 29.7\% & 591 & 16.9\% & 14.1\% \\ 
Variable Initialization & 476 & 12.3\% & 27.0\% & 297 & 8.5\% & 10.5\% \\ 
Data Type & 93 & 2.4\% & 56.8\% & 322 & 9.2\% & 36.4\% \\ 
Computation & 341 & 8.8\% & 24.9\% & 409 & 11.7\% & 21.2\% \\ 
\bottomrule   
\end{tabular}
\end{table}

This table illustrates that certain types of logic errors, such as loop conditions and data type issues, exhibit notably higher localization ratios, particularly in the S-Logic-Err dataset. These types of errors might be more distinctive or have clearer patterns that our model can detect effectively. In contrast, errors related to statement integrity and computation show lower localization ratios, indicating these may be more complex or involve subtler bugs that are harder for the model to pinpoint accurately. The high proportion of errors such as loop conditions and condition branches in both datasets reflects their commonality in programming, emphasizing the need for models like ours that can adeptly handle these frequent issues.

\section{Future Work}
In the field of programming education, merely localizing logic errors is insufficient. Future efforts should focus on enhancing Large Language Models to not only detect these errors but also generate more actionable feedback on logic errors, including suggested patches. This advancement would significantly improve learning outcomes by providing students with detailed guidance on how to correct their mistakes. Additionally, we recognize that the integration of multiple techniques, such as Graph Neural Networks, LSTM networks, and CodeBERT, increases the complexity and computational cost of our model. To address this issue, future research will explore model compression techniques, such as pruning and quantization, to reduce the computational footprint without sacrificing accuracy.

We also acknowledge the potential limitations of the graph attention mechanism, particularly in cases where the pseudocode-source code mapping is imperfect. To mitigate the impact of these imperfections, we plan to investigate alternative graph structures and incorporate confidence scores into the attention mechanism to handle uncertain mappings more effectively. Improving the accuracy of semantic alignment between pseudocode and source code, especially for abstract pseudocode, is another important direction for future work. We aim to refine the dynamic adjustment of alignment weights during training and further explore techniques to enhance semantic understanding in more complex or abstract cases. Finally, we plan to extend this logic error localization technique to support additional programming languages, such as Python and Java, which will increase the generalizability and applicability of the model in various educational and professional contexts. This extension is a priority in our ongoing research efforts.

\section{Conclusion}
This paper introduces a novel approach to logic error localization that combines semantic alignment with syntactic analysis, enhancing the identification of logic errors in student programs beyond traditional syntax error detection. Our method, validated through comprehensive experiments, outperforms current state-of-the-art tools in accuracy across multiple datasets. By integrating advanced techniques like bidirectional LSTM networks, graph attention mechanisms, and the semantic capabilities of CodeBERT, our framework not only identifies errors but also provides a foundation for future developments in automated program repair. This work lays the groundwork for further enhancements that could include generating corrective feedback and expanding to more programming languages, thereby improving both educational outcomes and programming proficiency. The code and dataset for independent evaluation are available at: https://github.com/Arrtourz/LogicerrorRepair.

\section{Acknowledgements}
This research was supported by the National Institute on Minority Health and Health Disparities (NIMHD) of the National Institutes of Health (NIH) under Award Number U54MD007595.

\bibliographystyle{splncs04}
\bibliography{b}

\end{document}